# Synthesis and Characterization of Copper Doped Zinc Oxide Thin Films for CO Gas Sensing

Sachin S Bharadwaj[a], Shivaraj B W[a*], H N Narasimha Murthy[a], M Krishna [a], Manjush Ganiger [a], Mohd Idris[a], Pundaleek Anawal[a], Vitthal Sangappa Angadi[a]

[a]Department of Mechanical Engineering, R V College of Engineering, Bangalore, 560059, India

**Abstract**

Objective of this work was to synthesize Copper doped Zinc Oxide (CZO) films and optimization of process parameters by varying molarity of zinc acetate dehydrate from 0.5 M to 1.0 M, concentration of copper acetate monohydrate from 1% to 5 % and annealing temperature from 200 ºC to 300 ºC to measure the sensitivity of CZO films for CO (Carbon Monoxide) gas. The concentration of CO gas was maintained at 5 ppm and operating temperature of 250 ºC was used for sensing. Analysis for sensitivity showed highest grading for parametric combination of 0.75 molarity, 3% copper concentration and 300 ºC annealing temperature with surface roughness of 3.90 nm and grain size of 256 nm. TEM image revealed the crystalline grain size was 5 nm. ANOVA showed that annealing temperature influenced the sensitivity by 69.06 % .

## 1. Introduction

With the growth of industries, simultaneously the working conditions in them as well as other such places have posed a lot of health hazards due to emission of harmful gases into the environment. The present air pollution testing techniques are expensive and time consuming. Owing to the low cost, easy installation, high sensitivity, quick response and reliability electronic thin film gas sensors have gained ground in present world scenario [1]. When materials are analyzed in the form of thin films their properties change to a large extent [2]. Thin film technology makes use of the fact that by varying the thickness various properties of materials can be changed. There exist few downsides in making use of undoped ZnO for optical devices as much wider band gap is required. Hence, it is very important to change the band gap of ZnO. This can be done by doping ZnO with impurities so addition of Copper into Zinc oxide is quite feasible and significant replacement of each other during the metastable thin film fabrication process and then resulting high solid solubility of CuO in ZnO for the thin film thus forming CZO alloys. The structural, electrical, optical properties of ZnO films are governed by deposition parameters like doping concentration, ageing time, spinning speed, post-annealing and number of layers of doped material.

The global gas sensors market was estimated at USD 1.98 billion in 2014, and is expected to grow at a Compound Annual Growth Rate of 5.1% from 2014 to 2020. The global gas sensors market is expected to reach a value of USD 2.32 billion by 2018. Carbon monoxide has 16.6% of total market share. Carbon monoxide (CO) is one among the most dangerous and harmful gases. It is extremely dangerous and highly toxic because it is colorless , odorless and it cannot be sensed by the human sense organs. CO gas exposure of 0.32% concentration for about 30 minutes can even be fatal for the human beings [3].

There are various methods that can be used in order to synthesize thin films such as sol-gel method, sputtering, spray pyrolysis [4] and laser deposition. Sol-gel method is widely used because of advantages like it is economical, ease of deposition, ease of controlling the composition of the films and controlling doping percentage [5-7]. Yogesh et al. [8] studied the properties of self aligned CZO films and found out that resistance decreased with doping of Copper in ZnO films. Copper doping of 1.5 % in ZnO, showed higher value of resistance (in the order of 109-1011 Ω) than that of 0.5 % and 1% , but undoped ZnO films showed much lower value of resistance. 1 wt% of copper doping gave higher sensitivity factor value for hydrogen.
The sensors showed operating temperature 220 °C for CO gas. Sayed et al. prepared [9] CZO thin films on glass substrates by sol–gel method and the spin coating technique. Copper doping with ratios less than 10 % molar ratio was chosen and incorporated into the spin coated ZnO thin films and observed significant modifications in both the structural as well as optical properties. The grain size gradually decreased as there was increase in Copper doping percentage from 0 to 9.8 %. Chow et al. [10] prepared Copper doped ZnO nanorods for hydrogen gas sensor. Copper doping was used to control structural and chemical properties. XRD studies showed that the slight difference in the lattice parameters of Copper doped ZnO was because of Copper atoms occupy various positions in the lattice which then form the defects. Tarek Saidani et al. [11] doped 5 % Copper into ZnO and dip coating technique was used for making films and deposited on glass substrates. Effect of post heating on various properties of films such as the structural, morphological and optical properties was studied. Post heating temperature was varied from 400°C to 550°C and the grain size also increased from 21.65 to 36.47 nm. Gong et al. [12] prepared nano structured crystalline copper doped zinc oxide thin film by c-sputtering and used of pure Zinc Oxide and Copper targets. The copper doped zinc oxide thin film exhibited a good response to CO gas at low concentration of 6 ppm and operating temperature 150 °C. Carbon monoxide sensing was operated at temperature 350 °C, which showed the response of 2.7 to 20. Spray pyrolysis studies on sensing properties [13] and even light enhanced sensing properties study has been done [14].

Research related to the effect on sensitivity by varying one factor at a time has been reported [11] and time consuming since it requires many experiments to co-relate the interaction of the operational parameters and hence resulting in an inefficient approach. In this research three parameters were selected i.e Copper doping concentration, Molarity and Annealing temperature has been varied to find out the influence on sensitivity of Copper doped Zinc Oxide films for CO gas. Design of Experiments was applied using MINITAB V16, a statistical software tool considering three factors, each at three variability levels. Synthesis of copper doped Zinc oxide thin films was done by sol-gel method. Prepared films were exposed to test gas (CO) and gas sensing test was conducted. Sensitivity of the films were calculated by making use of the data that was obtained during CO gas sensing test. Signal to Noise ratio (S/N ratio) analysis and ANOVA were carried out in order to analyze the effect of process parameters. Characterization of CZO thin film for which highest grading was obtained from grey relation analysis was carried out using AFM and SEM.

## 2. Experimental Details

### 2.1. Materials and Charaterization

Zinc acetate dihydrate [$Zn(CH_3COO)_2 \cdot 2H_2O$] was used as precursor material. Iso-propanol was used as a solvent material and Monoethanolamine (MEA) was used as the stabilizer. Copper acetate monohydrate $Cu(CH_3COO)_2 \cdot H_2O$ was used as the source of copper for the purpose of doping and glass was used as the substrate for deposition of thin films. The devices used were - Magnetic stirrer, Spin coater, Ultrasonicator (37 KHz 220 V, 500 ml ,250 Watts), Hot air oven and Furnace, Travelling Electron Microscope (TEM), Atomic Force Microscope (AFM), Scanning Electron Microscope (SEM) and Resistivity Multi-meter.

### 2.2 Preparation of CZO solutions

For the first experiment the mass of Zinc acetate dihyderate that was used is 2.173 g. The molarity of Zinc acetate dihyderate was kept constant at 0.5 M. The mass of Copper Acetate Monohydrate that was used is 0.02195g. Zinc acetate dihydrate and Copper Acetate Monohydrate were dissolved in Iso-Propanol of 20ml and then 2.15 ml of Monoethanolamine (MEA) was slowly added under magnetic stirring to prepare a solution, then the solution was stirred at 353K for 60 minutes using a magnetic stirrer to get a clear bluish sol. The weight ratio of MEA to zinc acetate was kept at 1:1 and dopant concentration was taken as 1%. The sol was kept for aging for 24 hours. Copper doped ZnO thin films were deposited by using spin coating on pre cleaned glass substrate at rotation speed of 2000 rpm. After spin coating, substrate was subjected to pre-heating of 150 °C for 10 minutes. This process of coating and pre-heating was carried for five layers. Then final film was annealed at 200 °C for 1 hour. Similar procedure was repeated for other experiments by changing the dopant concentration, molarity and annealing temperature.

### 2.3 Factors table and Characterization

Table 1 shows the Experimental table for the preparation of CZO fims and sensitivity was measured as response for each experiment.

Table 1 Design of Experiments for preparation of CZO films

| Experiment No. | Molarity (M) | Copper Concentration (%) | Annealing (°C) |
|---|---|---|---|
| 1 | 0.5 | 1 | 200 |
| 2 | 0.5 | 3 | 250 |
| 3 | 0.5 | 5 | 300 |
| 4 | 0.75 | 1 | 250 |
| 5 | 0.75 | 3 | 300 |
| 6 | 0.75 | 5 | 200 |
| 7 | 1.0 | 1 | 300 |
| 8 | 1.0 | 3 | 200 |
| 9 | 1.0 | 5 | 250 |

## 3. Results and Discussion
*3.1 Sensitivity of CZO films*

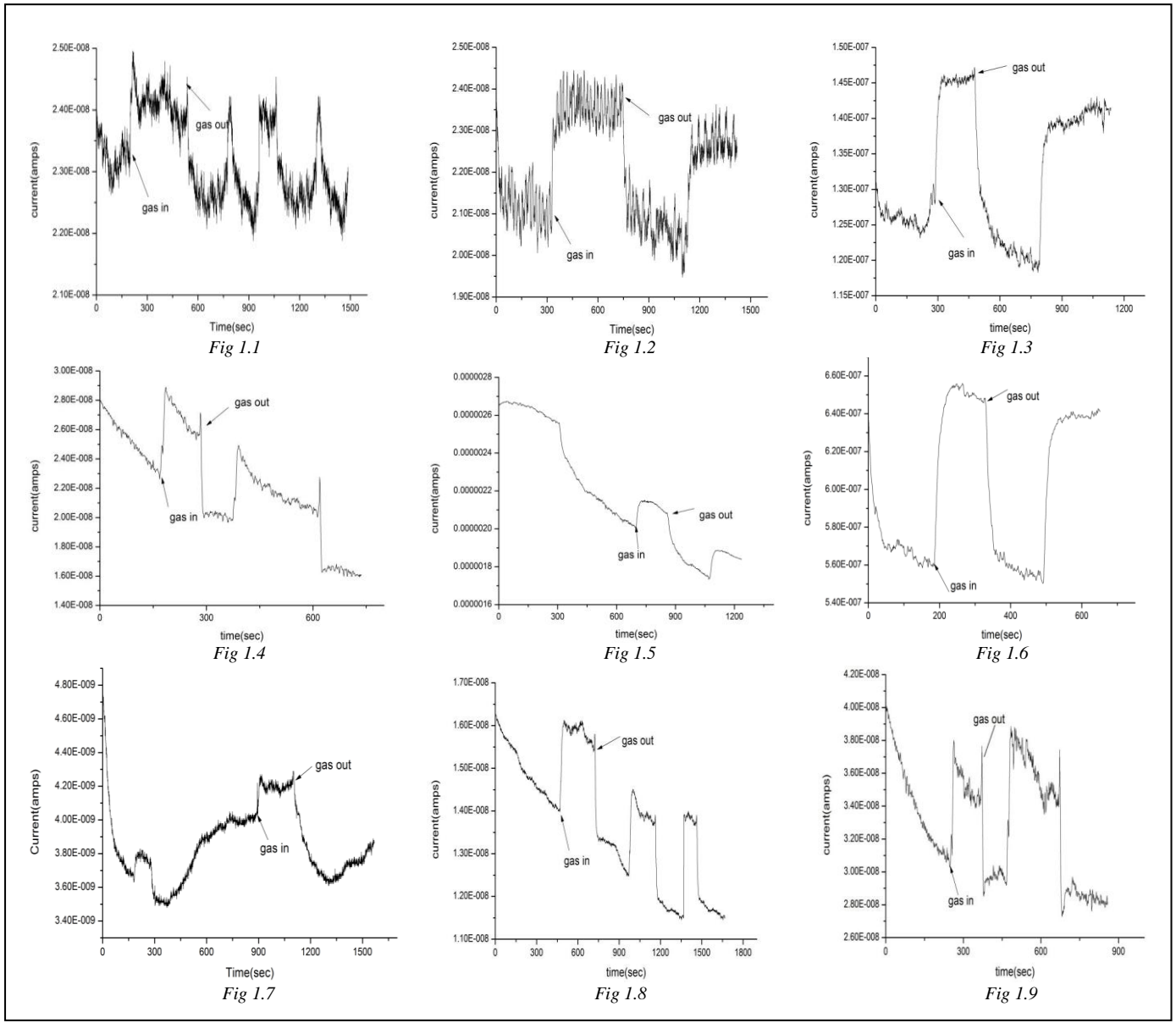

*Fig 1.1 to 1.9 shows the Current versus time graphs with varying morality, Copper concentration and post heating temperatures*

Fig. 1.1-1.9 shows results of sensitivity of CZO films. Sensitivity of 10.93 was observed in Experiment no. 5 with time period from 0 to 200 seconds represents the drop in current due to increase in the resistance of the film on exposure to the synthetic gas. On exposure to CO gas the amount of current flow increased from 5.7E-7 to 6.4E-7. The supply of CO gas stopped at 320 seconds and again at this point of time synthetic gas is sent into the chamber because of which the resistance increases. Equations (1), (2) and (3) shows the calculations for resistance and sensitivity.

Resistance in synthetic air atmosphere $\quad R = V/I$ .......................................... (1)

Sensitivity $\quad S = \Delta R/R = \left(R - \frac{Rg}{R}\right) * 100$ ..................... (2)

Resistance in CO atmosphere $\quad Rg = V/Ig$ ............................................ (3)

Table 2 Calculated values of sensitivity of CZO films

| Experiment No. | Sensitivity |
|---|---|
| 1 | 4.13 |
| 2 | 6.38 |
| 3 | 9.70 |
| 4 | 5.88 |
| 5 | 10.93 |
| 6 | 6.90 |
| 7 | 9.60 |
| 8 | 4.80 |
| 9 | 10.03 |

Table 2 shows that sensitivity results for all the experiments. Films produced had different sensitivity values due to different amount of doping concentration, Annealing temperature and Molarity of the films. In general, there was increase in the sensitivity values with increase in the percentage of the dopant concentration and also with the increase in the annealing temperature. Lowest value of sensitivity was obtained for the first sample which had lowest percentage of copper and also post heated at lowest temperature. The samples which had higher percentage of copper and those post heated at higher temperature showed higher sensitivity.

*3.2 Probability plots of sensitivity of CZO films*

Normal probability plot for sensitivity is shown in Fig 2. From the experiment results, it was found that sensitivity results to be equally distributed along the trend line of a normal probability plot which indicates that the synthesis is carried out at stable conditions and hence was not influenced by any extraneous factors. Since all the values lie within the lines of the probability plot no experiment needs to be carried out once again.

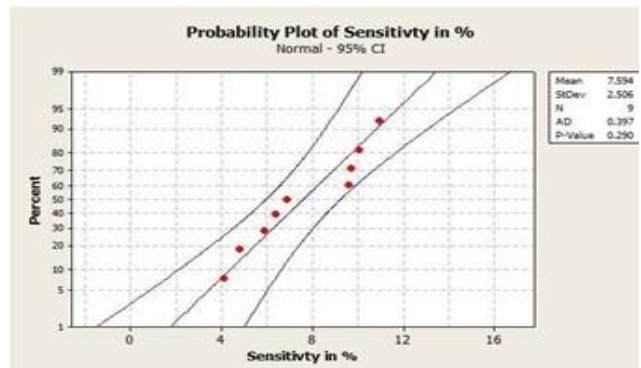

*Fig 2 Probability plot for sensitivity*

*3.3 S/N ratio for Sensitivity*

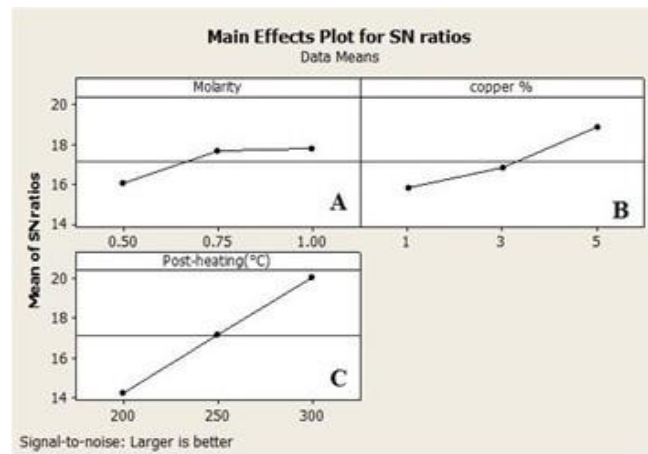

*Fig 3 S/N Ratio for sensitivity*

Fig. 3 shows the effect of Molarity, copper concentration and annealing temperature on CZO films. Sensitivity of Larger the better was considered for calculating the S/N ratio. Highest S/N ratio was obtained for parametric combination of 1M, 5 % copper doping and increased the molarity and annealing temperature of 300˚C.

*3.4 Interactions plot for Sensitivity*

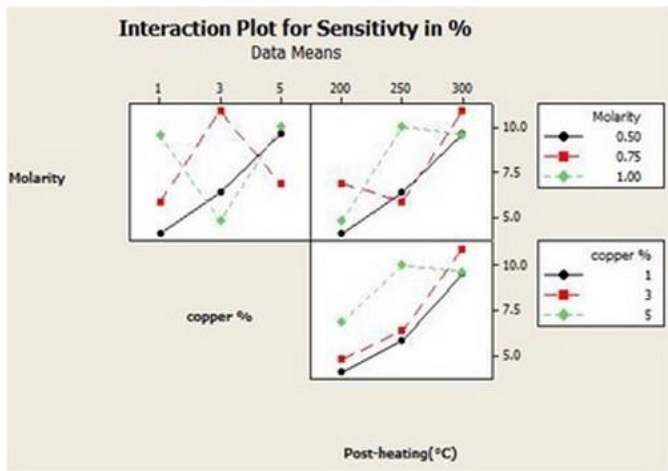

*Fig 4 Interaction plot for sensitivity*

Fig.4 shows that for 0.75 M of solution, 3 % of Copper concentration higher sensitivity is obtained. Lowest value is for 0.5 M solution and 1 percentage of copper. 300 °C annealing temperatureand 0.75 M solution gives higher value of sensitivity and 0.5 M solution and 200°C annealing gives the lowest value for sensitivity. Also 3% of copper and 300 °C temperature gives a higher value of sensitivity and 1% of copper and 200°C annealing give lowest value for sensitivity. Sensitivity is larger the better for CZO films as higher value of sensitivity[14] is desired.

*3.5 ANOVA for sensitivity of CZO films*

Table 3 shows the ANOVA results for the sensitivity of CZO film to study the contribution of each parameter. It can be observed that the molarity (P=6.76%), copper doping (P = 16.8%) and annealing temperature (P = 69.06 %) showed significant influence on the sensitivity of film [13].

*Table 3 ANOVA result for sensitivity*

| Factors | DF | SS | MS | F | P (%) | $F_{0.05}$ |
|---|---|---|---|---|---|---|
| Molarity (M) | 2 | 3.397 | 1.699 | 0.92 | 6.76 | 3 |
| Copper (%) | 2 | 8.440 | 4.220 | 2.28 | 16.80 | 3 |
| Annealing Temperature (°C) | 2 | 34.682 | 170341 | 9.37 | 69.06 | 3 |
| Error | 2 | 3.702 | 1.851 | | 7.37 | |
| Total | 8 | 50.221 | | | 100.00 | |

DF = Degree of freedom; SS = Sum of square; MS = Mean square; P = Percentage of contribution, F= F-test, $F_{0.05}$= 95% confidence band.

*3.6 Characterization*

Fig. 5.shows the TEM image for highest sensitivity of CZO films with crystalline grain size was 5 nm. Fig. 6 shows the scanning electron microscope image obtained for the film, with a grain size of was 256 nm. Fig. 7 shows 3D and Fig. 8 shows 2D AFM images of CZO films with a surface roughness of 3.90 nm. Larger the value of roughness, better it is for gas sensing as more number of oxygen molecules get adsorbed on the surface of the film [12]. Surface roughness increases with the increase in the copper percentage and increase in the annealing temperature [5] and also sensitivity of the films will increase with doping concentration and annealing temperature based on the roughness of the films.

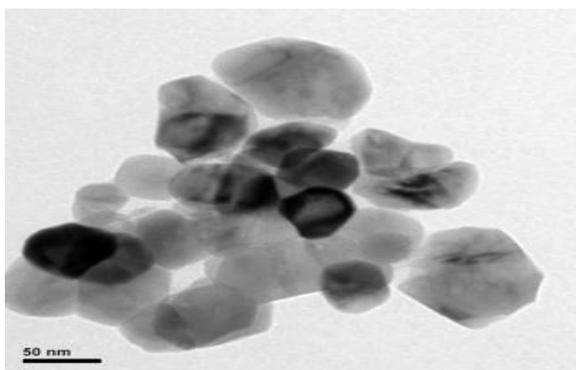
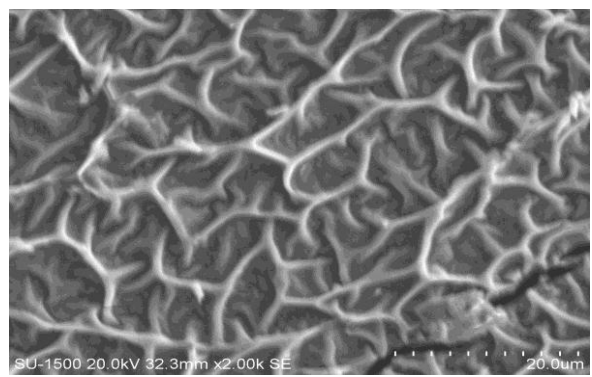

*Fig 5 TEM image for CZO films*        *Fig 6 SEM image for CZO films*

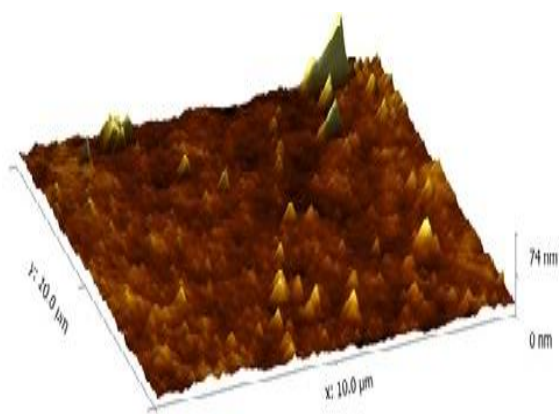
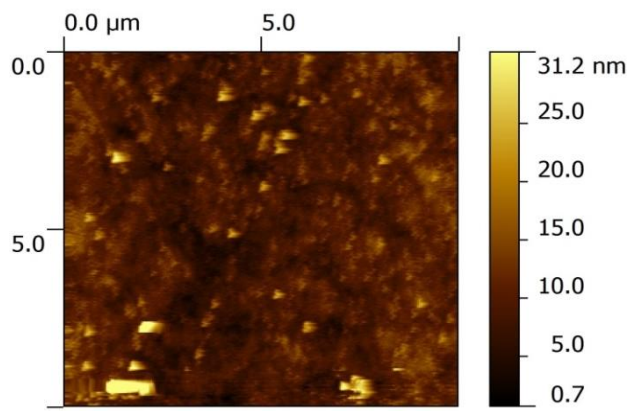

*Fig 7  3D view of AFM image of CZO Films*        *Fig 8  2D view of AFM image of CZO Film.*

## 4. Conclusion

Sensitivity measurement for all 9 thin films was carried out and it is found that sensitivity was maximum for 5[th] experiment sample, having combination of 0.75 M, 3 % copper and 300ºC annealing temperature. ANOVA results showed that annealing temperature showed the significant influence on the sensitivity of CZO film .TEM image revealed the grain size was 5 nm CZO films. SEM results showed that the CZO films have a grain size of was 256 nm with dendrite structure. AFM images of CZO films with a surface roughness of 3.90 nm. Thus the results of the experiments were optimized to achieve the best sensitivity for given CZO gas sensor. Thus this would truly improve the sensing characteristics of the given gas sensor under the conducted experimental conditions.

**Acknowledgements**

Financial support for the present research was provided by TEQIP-II Sub – Comp: 1.2.1 RVCE